\documentclass[journal]{IEEEtran}
\usepackage{cite}

\ifCLASSINFOpdf
\else
\fi
\usepackage{graphicx}

\graphicspath{{Figures/}}
\usepackage{caption}
\usepackage{capt-of}
\usepackage{balance}
\usepackage{subfigure}
\usepackage{fixltx2e}
\usepackage{setspace}
\usepackage{amsmath}
\usepackage{array}
\usepackage{hyperref}
\usepackage{breqn}

\begin{document}

	\title{A Robust and Accurate Deep Learning based Pattern Recognition Framework for Upper Limb Prosthesis using sEMG }
	%
	%
	%
	\author{Sidharth~Pancholi,~\IEEEmembership{Member,~IEEE},  Amit M.~Joshi,~\IEEEmembership{Senior Member,~IEEE}, and Deepak Joshi,~\IEEEmembership{Senior Member,~IEEE}
		\thanks{Sidharth Pancholi was with the Department of Electronics and Communication Engineering, Malaviya National Institute of Technology, Jaipur, Rajasthan,India e-mail: (s.pancholi@ieee.org)}
		
		\thanks{Amit M. Joshi was with the Department of Electronics and Communication Engineering, Malaviya National Institute of Technology, Jaipur, Rajasthan,India e-mail: (amjoshi.ece@mnit.ac.in)}
		
		\thanks{Deepak Joshi is with the Centre for Biomedical Engineering, Indian Institute of Technology, Delhi, India (e-mail: Deepak.Joshi@cbme.iitd.ac.in)}
		\thanks{Manuscript received April 19, 2005; revised August 26, 2015.}}

	\maketitle

	\begin{abstract}
		In EMG based pattern recognition (EMG-PR), deep learning-based techniques have become more prominent for their self-regulating capability to extract discriminant features from large data-sets. Moreover, the performance of traditional machine learning-based methods show limitation to categorize over a certain number of classes and degrades over a period of time. In this paper, an accurate, robust, and fast convolutional neural network-based framework for EMG pattern identification is presented. To assess the performance of the proposed system, five publicly available and benchmark data-sets of upper limb activities were used. This data-set contains 49 to 52 upper limb motions (NinaPro DB1, NinaPro DB2, and NinaPro DB3), Data with force variation, and data with arm position variation for intact and amputated subjects. The classification accuracies of 91.11\% (53 classes), 89.45\% (49 classes),  81.67\% (49 classes of amputees), 95.67\% (6 classes with force variation), and 99.11\% (8 classes with arm position variation) have been observed during the testing and validation. The performance of the proposed system is compared with the state of art techniques in the literature. The findings demonstrate that classification accuracy and time complexity have improved significantly. For signal pre-processing and deep learning techniques, Keras which is a high-level API for TensorFlow to build profound learning models has been utilized. The proposed method has been tested on the Intel Core i7 3.5GHz, 7th Gen CPU with 8GB DDR4 RAM.
	\end{abstract}
	

	\IEEEpeerreviewmaketitle

	\section{Introduction}

\IEEEPARstart{S}{urface} Electromyography or sEMG is a significant physiological parameter that is used in the characterization of neuromuscular pathologies, muscle computer interfaces (MCI) such as bio robotics and human motor control based applications \cite{atzori2012building, Pancholi2019}. In general, sEMG signals are used in EMG-PR based systems for pattern recognition and subsequently supervised learning is utilized for pattern classification. 
	
	\begin{figure}[htbp]
		\includegraphics[width=8.5cm,height=6cm]{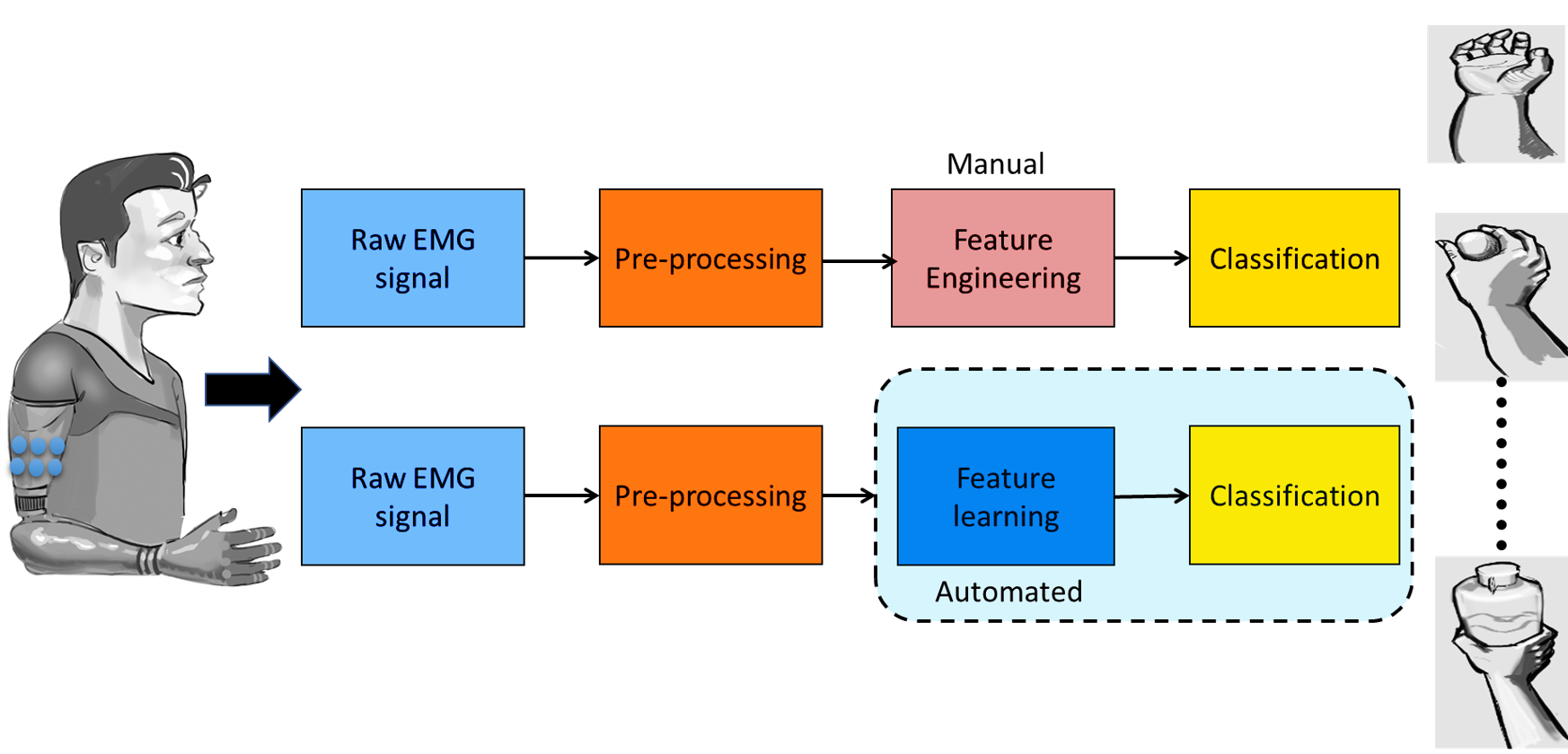}
		\centering
		\caption{Functional diagram of deep learning based EMG-PR system for upper limb motion classification}
		\label{theam}
	\end{figure}

It is considered that employing an instantaneous EMG signal to a pattern recognition model is not effective for pattern classification. Classical EMG motion recognition technique requires processing steps such as: (i) Data acquisition and pre-processing, (ii) feature extraction, (iii) dimensionality reduction, and (iv) motion identification or classification\cite{lorrain2011influence}, \cite{pancholi2018portable},  \cite{pancholi2016development}.  Due to non-stationary and stochastic property of the EMG, the instantaneous measures of the signal are inappropriate for the training of the traditional machine learning algorithm \cite{atzori2016deep}. Therefore, in the feature extraction, a short time window (typically between 100-250ms) is applied to produce additional stability and enhance information concentration \cite{Smith2011}. This step is required before a classification algorithm to compute the output on raw EMG signal. Several time, frequency, and time-frequency domains have been investigated for feature extraction in recent years. One of the renowned feature extraction techniques consists a set of MAV (mean absolute value), SSC (number of times slope sign change), ZC (zero crossings), WL (waveform length) named as Hudgins time-domain feature set \cite{tkach2010study}, \cite{adewuyi2016analysis}. In recent years, short-time Fourier Transform (STFT) and continuous Wavelet transform (CWT) have been reported for pattern detection of sEMG data. \cite{guo2017toward}. Time derivative moment and fused wavelet packet transform-based features have been also proposed and observed a significant enhancement in pattern recognition performance \cite{pancholi_improve}, \cite{pancholi2019time}.  However, handcrafted features have played a significant role in EMG based pattern classification so far, deep learning has recently begun to showcase greater achievements than hand-engineered features \cite{zhang2019real}, \cite{ding2018semg}. Furthermore, the existing system for grasping motions only satisfy to a certain extent human-centric activities, These include the necessity for a large number of functions, faster reaction/execution times, and control system intuitiveness. Fig. \ref{theam} depicts the functional diagram of the deep learning-based motion classification technique.\par
	
Deep learning algorithms have grown more prevalent in recent years because they are unrivalled in their capacity to automatically learn discriminate features from large volumes of data. The basic goal of deep learning is to shift the emphasis from human feature extraction to automated feature learning \cite{pancholi2021source}. However, neural network-based architecture has been widely used for EMG-based motion recognition. Deep learning algorithms have had a significant influence on pattern recognition employing high-end computer workstations in the Big Data era. Convolutional neural networks (CNNs) are a type of deep neural network that has garnered a lot of traction in the field of pattern recognition. End-to-end learning models can be solved using CNN-based architecture. The spectral representation of EMG signals, such as Short-time Fourier transform (STFT) and Wavelet transform-based images, have been investigated to be fed to these neural networks. \cite{alaskar2018deep, Kumar2003}. Atzori et al. \cite{atzori2016deep}  has explored CNN with different classification techniques on NinaPro Data sets. The average classification accuracies of 66.6\% and 60.30\% have been achieved on NinaPro sub-database 1 and 2. In the research \cite{cote2019deep}, the classification accuracy of 68.98\% has been achieved for the NinaPro DB5 dataset (18 hands/wrist gestures), and this experiment was conducted over 10 subjects on a single Myo Armband. The main factor for accurate recognition when dealing with deep learning algorithms is the amount of training data available. The deep learning-based methods require the signal to image conversion \cite{lorrain2011influence}, the large amount of data, and a long time for the training\cite{geng2016gesture}. The translation deep learning technique to deploy for prosthesis application is still challenging. This requires large amount of data and training time. To address these problems, the goal of this work is to shorten training time and improve the performance of the CNN-based architecture for EMG-PR applications. The following are the primary contributions of this work: I a novel Deep Learning-based Pattern Recognition (DLPR) framework that uses a CNN to improve the performance of sEMG-based gesture recognition, (ii) data classification was done on five open source challenging and heterogeneous data sets that included both intact and amputated participants. (iii) analysis of classification accuracy in relation to DASH (disability of the arm, shoulder, and hand), (iv) training time of the suggested mode for all data-sets, and (v) comparison of the proposed work with state-of-the-art methodologies.

This paper is organized as follows. The information about experimental data has been discussed in Section II. Section III presents the proposed DLPR architecture for EMG-PR based bio-robotics and prosthesis. The result is emphasized in section IV. Section V covers the discussion about the result. The conclusion of this work is covered in Section VI.\\ 
	
	\begin{table}[htbp]
		\caption{Datasets information of heathy subjects (NinaPro DB1 and NinaPro DB2)}
		\begin{tabular}{llll}
			\hline \hline
			Parameters          & NinaPro DB1 & NinaPro DB2 \\ \hline \hline
			Number of subjects  & 27          & 40          \\
			Considered subjects & 10          & 10          \\
			Avg. years          & 28$\pm$4.6     & 28$\pm$3.1\\
			Avg. height (cm)         & 173.1$\pm$7.6  & 17$\pm$11 \\
			Avg. weight (kg)         & 68.6$\pm$12.0  & 69.9$\pm$13.8\\ \hline
		\end{tabular}
		\label{db12}
	\end{table}
	
	\begin{table*}[]
		\centering
		\caption{NinaPro DB3: Total 5 transradial amputees}
		\begin{tabular}{llllllllll}
			\hline \hline
			\begin{tabular}[c]{@{}l@{}}Subject\\   ID\end{tabular} & Hand       & Age & \begin{tabular}[c]{@{}l@{}}Height\\  (cm)\end{tabular} & Weight (kg) & \begin{tabular}[c]{@{}l@{}}Remaining\\   forearm\\     (\%)\end{tabular} & \begin{tabular}[c]{@{}l@{}}Years Passed \\ by the \\ amputation\end{tabular} & \begin{tabular}[c]{@{}l@{}}Amputation \\ cause\end{tabular} & \begin{tabular}[c]{@{}l@{}}Phantom\\ Limb\\ Sensation\\ Intensity\end{tabular} & \begin{tabular}[c]{@{}l@{}}DASH\\ score\end{tabular} \\ \hline \hline
			2                                              & Left   & 35  & 183                                                    & 81     & 70                                                                       & 6                                                                            & Accident                                                    & 5                                                                              & 15.18                                                \\
			4                                              & Right  & 34  & 166                                                    & 68     & 40                                                                       & 1                                                                            & Accident                                                    & 1                                                                              & 86.67                                                \\
			8                                              & Right & 33  & 175                                                    & 85     & 50                                                                       & 5                                                                            & Accident                                                    & 2                                                                              & 33.33                                                \\
			9                                              & Right & 44  & 180                                                    & 95     & 90                                                                       & 14                                                                           & Accident                                                    & 5                                                                              & 3.3                                                  \\
			11                                             & Right & 45  & 183                                                    & 75     & 90                                                                       & 5                                                                            & Cancer                                                      & 4                                                                              & 12.5                                                 \\ \hline
		\end{tabular}
		\label{db3}
	\end{table*}

	\section{EXPERIMENTAL DATA}
	
The performance of the proposed DLPR architecture has been validated on five publicly accessible,  scientific, and benchmark datasets NinaPro dB1, NinaPro dB2, and NinaPro dB3\cite{atzori2012building} (\url{http://ninapro.hevs.ch/}), upper limb motions with force  variations\cite{al2015improving} (DB4), and variant limb positions\cite{khushaba2014towards} (DB5) (\url{https://www.rami-khushaba.com/electromyogram-emg-repository.html}). The demographic information of data-set-1 and data-set-2 can be found in Table \ref{db12}. In the NinaPro DB1, an EMG signal of a total of 27 subjects has been recorded with otto Bock 13E200 electrodes and a filter of 90 to 450 Hz has been applied. This data-set contains 52 upper-limb motions and these motions have been divided into 3 categories: (i) 12 Basic movements of fingers, (ii) 8 Isometric, isotonic hand configuration with 9 basic wrist movements and (iii) 23 Grasping and functional movements.

\begin{figure}[htbp]
\includegraphics[width=8.5cm,height=6cm]{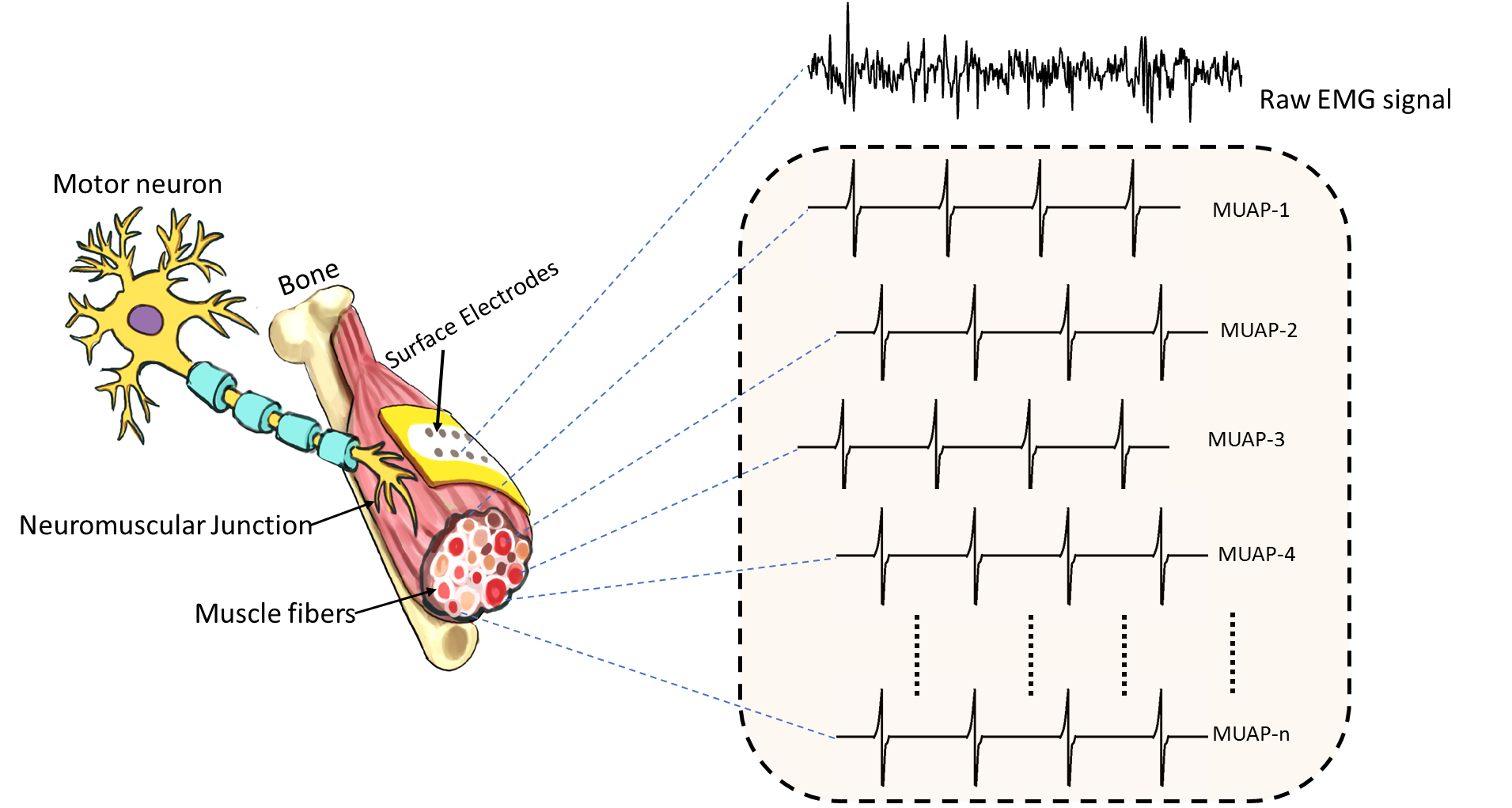}
\centering
\caption{
EMG signal formation: The resultant EMG signal is the product of each motor neuron's activity, which triggers the production of action potentials in the muscle fibre. The superimposed action potential of the fibres (motor unit action potential) innervated by each motor neuron is acquired at the skin's surface, and the combined activity of all active motor units produces the EMG.}
		\label{EMG}
	\end{figure}

EMG data of 40 subjects have been acquired in the NinaPro DB2 with Delsys Trigno wireless system and a band-pass filter of 20 to 450 Hz has been utilized. This data-set was acquired for 17 fingers and wrist movement, 23 grasping and functional motions, and 9 motions for different force pattern applied on fingers. NinaPro DB3 has also listed into 3  categories same as NinaPro DB1 for transradial amputees. In this work,  10 subjects have been taken into consideration from NinaPro DB1 and NinaPro DB2. Subsequently, 5 transradial amputees have been selected from the NinaPro DB3. The details of the selected subjects have been presented in Table \ref{EMG}.  In the DB4, a total of 9 subjects comprises two congenital and 7 traumatic have participated. The movements are: (i) Thumb flexion, (ii) Index flexion, (iii) Fine pinch, (iv) Tripod grip, (v) Hook grip (hook or snap), (vi) Spherical grip power. Each subject is requested to use his intact-hand to imagine the required movement with the selected force level. Moreover, they utilized Visual Feedback (VF) from the system screen to see the EMG signals of all channels to make it helpful for them to produce the desired force. For each of the 6 grip patterns, the amputees produced 3 force levels: low, medium and high. In every phase of different force levels, 5 to 8 times were acquired for each amputee where each activity had a holding phase of 8-12 seconds. Further, in DB5, the same movements were performed at different limb positions for different EMG patterns. Total 11 subjects including 9 males and 2 females, of  20 to 37 years of age were selected to carry out 8 movements. A set of 8 hand motions, including  (i) wrist flexions, (ii) wrist extensions, (iii) pronation, (iv) supination, (v) power grip, (vi) pinch grip, (vii) open hand and (viii), rest, were performed at various limb locations by each subject.

	\begin{figure*}[h]
		\includegraphics[width=18cm,height=10cm]{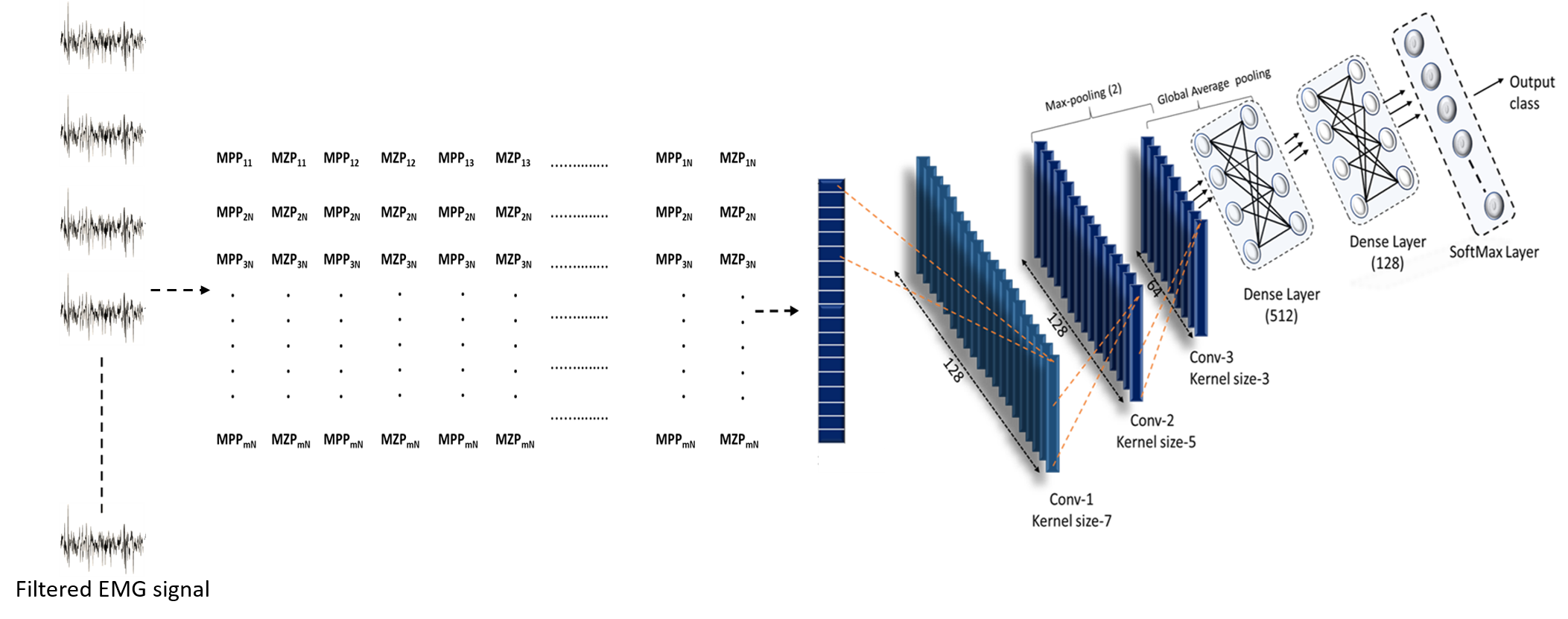}
		\centering
		\caption{Proposed DLPR architecture for EMG-PR based applications}
		\label{proposed}
	\end{figure*}
	
	\section{PROPOSED DLPR FRAMEWORK FOR EMG-PR}
	
	Raw myoelectric signal for EMG-PR application is ineffective because the raw EMG signal is non-linear, non-stationary, and stochastic in nature \cite{farina2002motor}. This property of EMG signal is because of continuous variation in motor unit recruitment and arbitrary manner in which these motor unit action potentials are superimposed as shown in Fig. \ref{EMG}.   
	
	The time differential property of the Fourier Transform suggests that the $n^{th}$ derivative of a signal can be determined by multiplying the frequency spectrum of a signal by $k$ raised to the power $n$. It is specified in the equation (\ref{eq1}).

	\begin{equation}
	F\left [ \triangle ^n \sigma (j)\right ]=k^n\chi \left [ k \right ]
	\label{eq1}
	\end{equation}
	Where $\sigma (j)$ signal in time domain, $\triangle ^n$ represents the $n^{th}$ derivative of signal, and $\chi \left [ k \right ]$ is the frequency transform of the signal. 
	
	The power spectral moments are used to pre-process the signal before feeding input to the neural network. The moment $\mu$ of $n^{th}$ order for window length $T$ is defined as in equation (\ref{eq2}).
	
	\begin{equation}
	\mu_n=\sqrt{\sum_{k=0}^{T-1}k^n\chi\left [  k\right ]}
	\label{eq2}
	\end{equation}
	
The IOS (integral square) of the segmented signal can be defined as its power, as the parseval ($\eta$) theorem suggests as presented in the equation (\ref{parseval}). 
	
	\begin{equation}
	\sum_{j=0}^{T-1}(\sigma\left [ j \right ])^2=\frac{1}{T}\sum_{k=0}^{T-1}\left |  \chi  \left [k  \right ].\chi  \left [k  \right ]^{*}\right |=\sum_{k=0}^{T-1}\eta \left [ k \right ] 
	\label{parseval}
	\end{equation}
	As per the equations (\ref{eq1}), (\ref{eq2}), and (\ref{parseval}) the $0^{th}$, $2^{nd}$, and $4^{th}$ order moments are defined as follows:

	\begin{align}
	\mu_0=\sqrt{\sum_{j=0}^{T-1}(\sigma[j])^{2}}
	\label{zero}
	\end{align}

	\begin{align}
	\mu_2=\sqrt{\sum_{j=0}^{T-1}(\Delta \sigma[j])^{2}}
	\label{three}
	\end{align}

In the same way as described previously, the fourth order moment is depicted in equation (\ref{four}).

	\begin{align}
	\mu_4=\sqrt{\sum_{k=0}^{T-1}k^{4}\chi[k]}=\sqrt{\sum_{j=0}^{T-1}(\Delta^{2} \sigma[j])^{2}}
	\label{four}
	\end{align}
	
The NPs (number of peaks) and ZCs (zero crossings) of a stochastic process can be calculated using their spectral moments \cite{dirlik1985application}, \cite{pancholi2019time}. This is defined in equations (\ref{NPs}) and (\ref{ZCs}). 
	
	\begin{align}
	NPs = \sqrt{\frac{\mu_4}{\mu_2}} 
	\label{NPs}
	\end{align}

	\begin{align}
	ZCs = \sqrt{\frac{\mu_2}{\mu_0}} 
	\label{ZCs}
	\end{align}
	
The square version of NPs and ZCs has been used to simplify equations (\ref{NPs}) and (\ref{ZCs}), as illustrated in equations below.
	
	\begin{align}
	NPs={\frac{\mu_4}{\mu_2}}=\Psi
	\label{NPs1}
	\end{align}
	
	\begin{align}
	ZCs={\frac{\mu_2}{\mu_0}}=\Phi
	\label{ZCs1}
	\end{align}

The sEMG signal is passed via two functions in the recommended architecture: I MPP ($psi$), which is the multiplication of peaks and power, and (ii) MZP ($phi$), which is the multiplication of zero crossings and power. This phase stabilises the EMG signal and reduces the size of the training data set while maintaining neuronal information. The peaks of these parameters include frequency information, whereas the zero order moment includes time information without any time to frequency transformation.
	
	\begin{align}
	MPP  = {\mu_0}*{\Psi }
	\label{NPs2}
	\end{align}

	\begin{align}
	MZP = {\mu_0}*{\Phi }
	\label{ZCs2}
	\end{align}

As shown in Fig.\ref{proposed} the suggested CNN structure is made up of six blocks: three convolution layers, two fully connected layers, and one softmax layer for classification. The initial convolution layers, which comprise 128 filters with a kernel size of 7, are used to feed pre-processed data. The output of the first convolution layer is 18 x 128. A 14x128 frame is created by the second convolution layer. Following that, the design has been updated to accommodate a size 2 max pool, resulting in a 7x128 output. The third convolution layer, which comprised of 64 filters with a kernel size of 3, has also been employed. The output of the third convolution layer is 5x64. After that, global average pooling is employed between the third and fourth fully connected (FC) layers of size 512. The softmax layer is then added, followed by an FC of 128. The Adam optimization strategy has been used to apply batch normalisation and a ReLU non-linearity after each block. The mathematical formulation for layer-1 can be written as follows in equation (\ref{layer1}).

	\begin{equation}
	\alpha _{m}^{l}(j)=b^{l}(j)+\sum_{i=1}^{i\leq 128}\left ( I_{i,j} w_1\left ( l,m,i \right )+I_{i,j+1}w_2(l,m,i)\right )
	\label{layer1}
	\end{equation}
	where $\alpha _{m}^{l}(j)$ belongs to neuron $j$  of map m in layer $l$. It also shows the multiplication between the weighted values and input  neurons.The filter used in the neural network is $w_k\left ( l,m,i \right)$. layer, map, channel/kernel and the position are defined by  $l$, $m$, $i$ and $k$ in the kernel respectively. The bias in the layer-1 for the neuron j is indicated by $b^{l}(j)$. For layer-2 and layer-3 the mathematical expression is defined as equation (\ref{layer2}).
	
	\begin{equation}
	\alpha _{m}^{l}(j)=b^{l}(j)+\sum_{i=1}^{i\leq N_{kernel}^{l}}(M_{i,j}w_{l}(l,m,i)+M_{i,j+1}^{l-1}w_{2}(l,m,i))
	\label{layer2}
	\end{equation}
	The number of kernels are  denoted by $ N_{kernel}^{l}$ in layer $l$. $M_{i,j}$ belongs to the components i, j (neuron j in-kernel i) of the feature maps produced following convolution from the preceding layers. Finally, 2 fully connected layers (FCs) are used and the softmax layer of size 23 is employed.The classification accuracy metric has been considered for the performance evaluation as denoted in equation \ref{accu} and each data-set has been divided into two parts: (i) training data (60\% of the total data), (ii) testing data (40\% of the total data). 
	
\begin{dmath} 
Classification\,accuracy=\frac{Total\,number\,of\,correct\,predictions} {Total\,tested\,samples}
\label{accu}
\end{dmath}

	\begin{figure*}[htbp]
		\centering
		\subfigure[]{\includegraphics[width=0.45\textwidth,height=0.25\textheight]{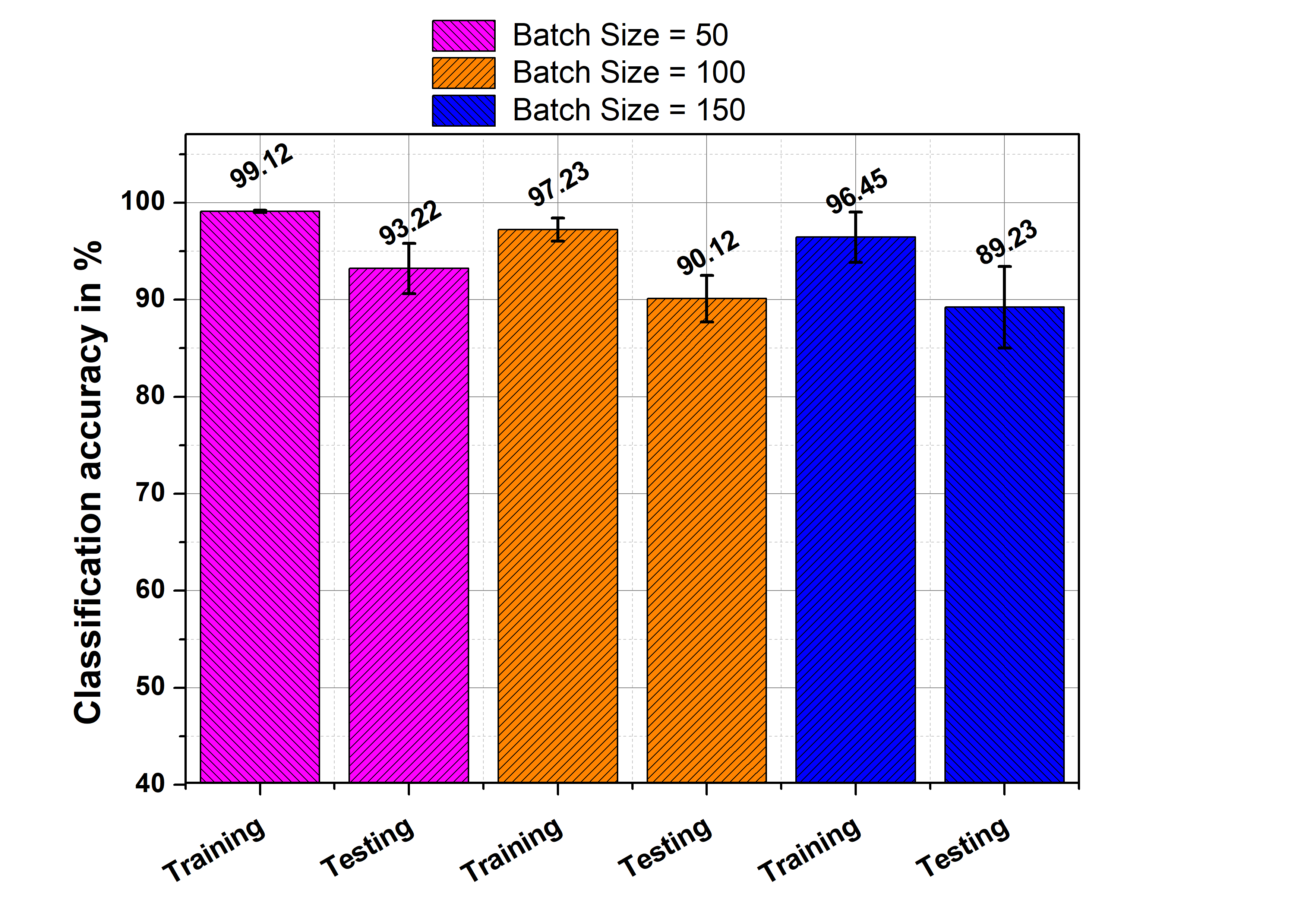}}
		\subfigure[]{\includegraphics[width=0.45\textwidth,height=0.25\textheight]{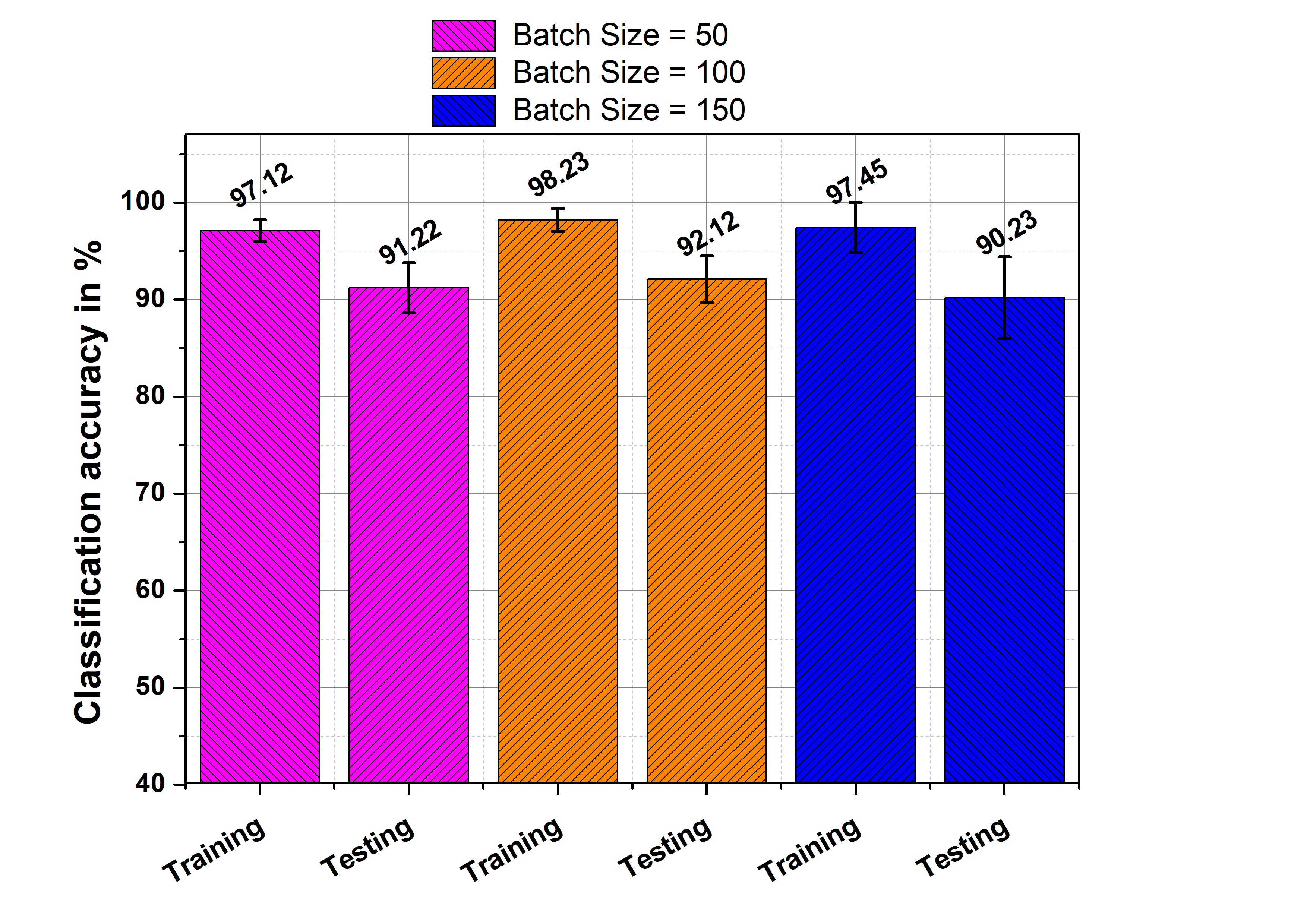}}
		\subfigure[]{\includegraphics[width=0.45\textwidth,height=0.25\textheight]{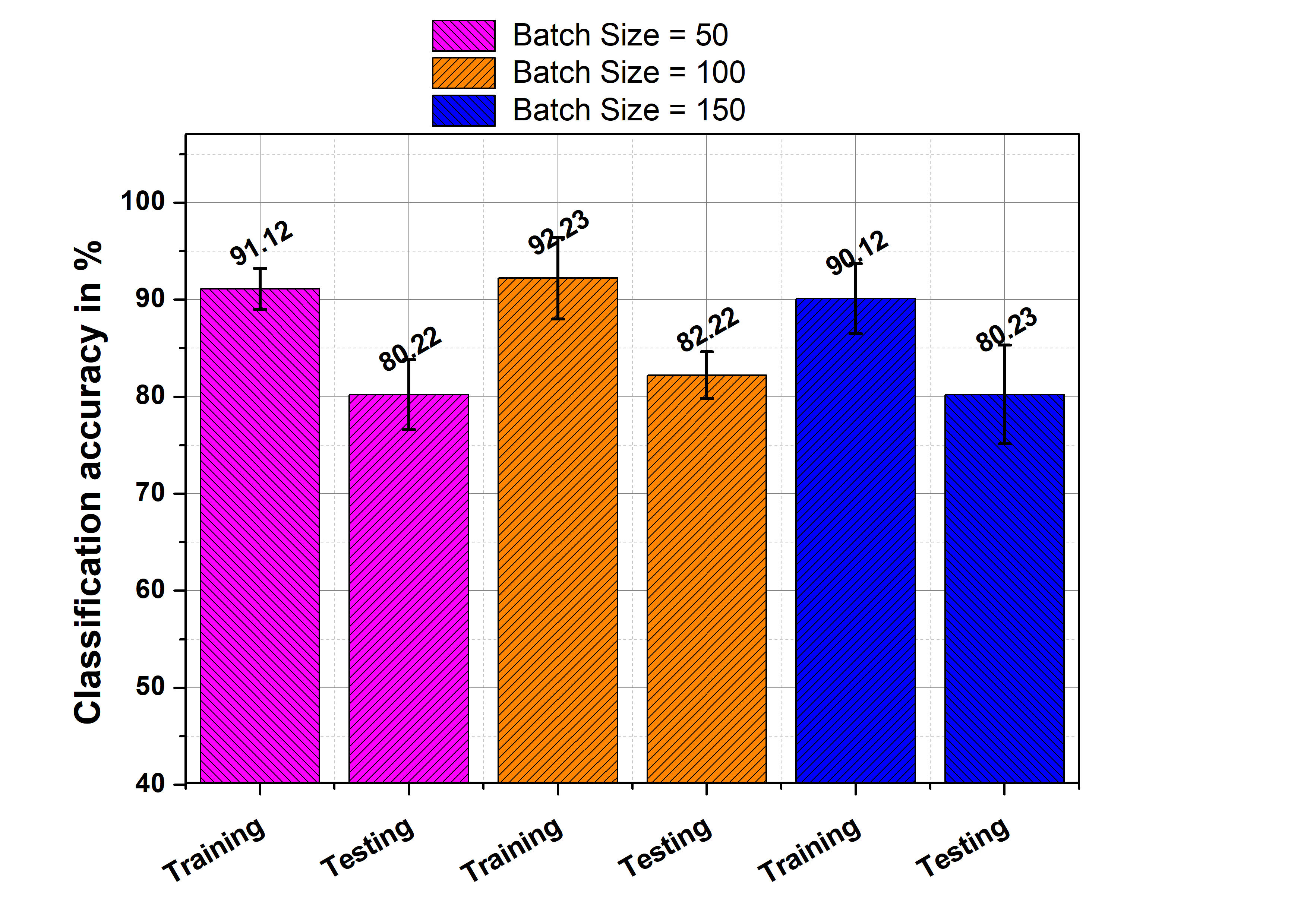}}
		\subfigure[]{\includegraphics[width=0.45\textwidth,height=0.25\textheight]{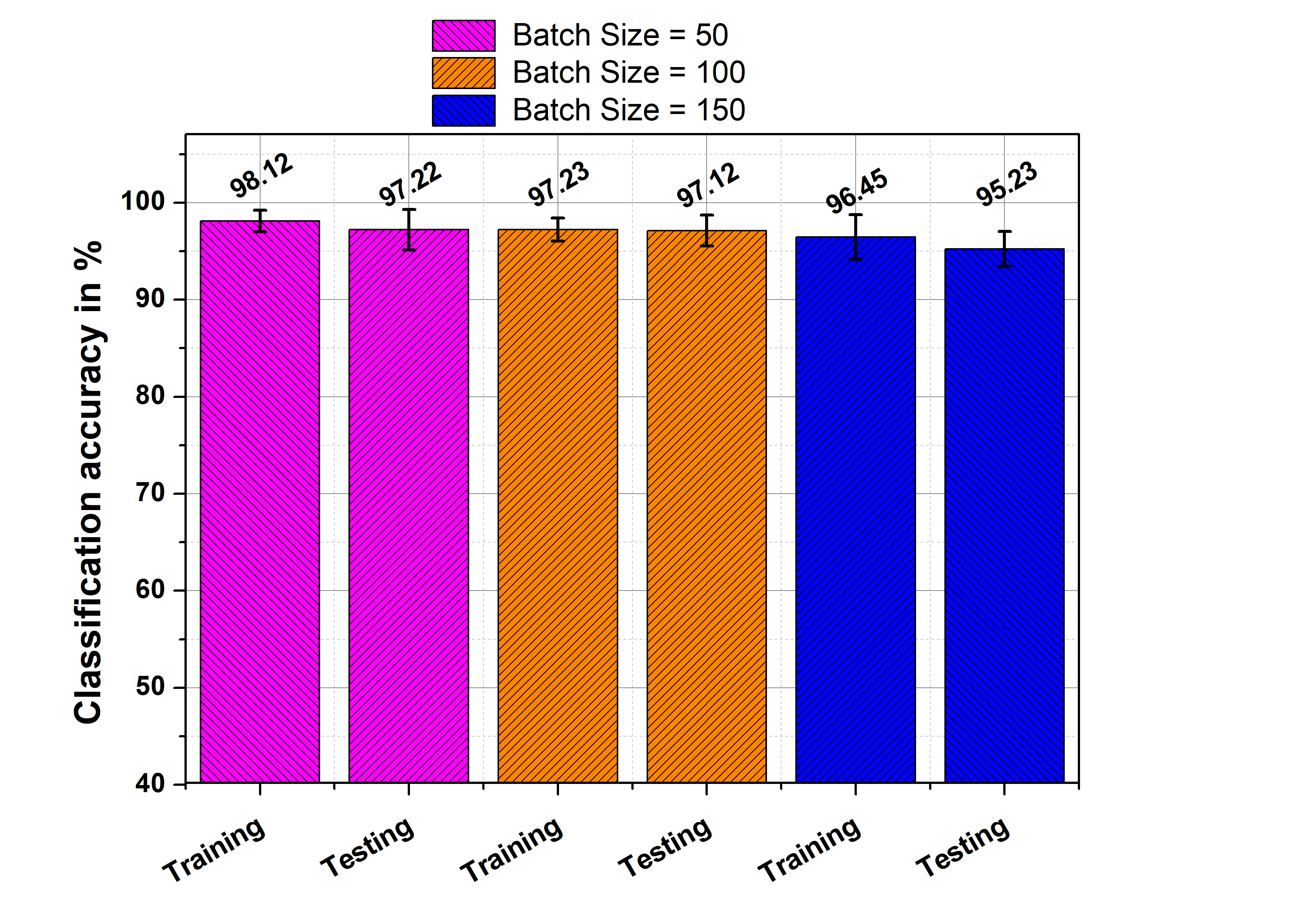}}
		\subfigure[]{\includegraphics[width=0.45\textwidth,height=0.25\textheight]{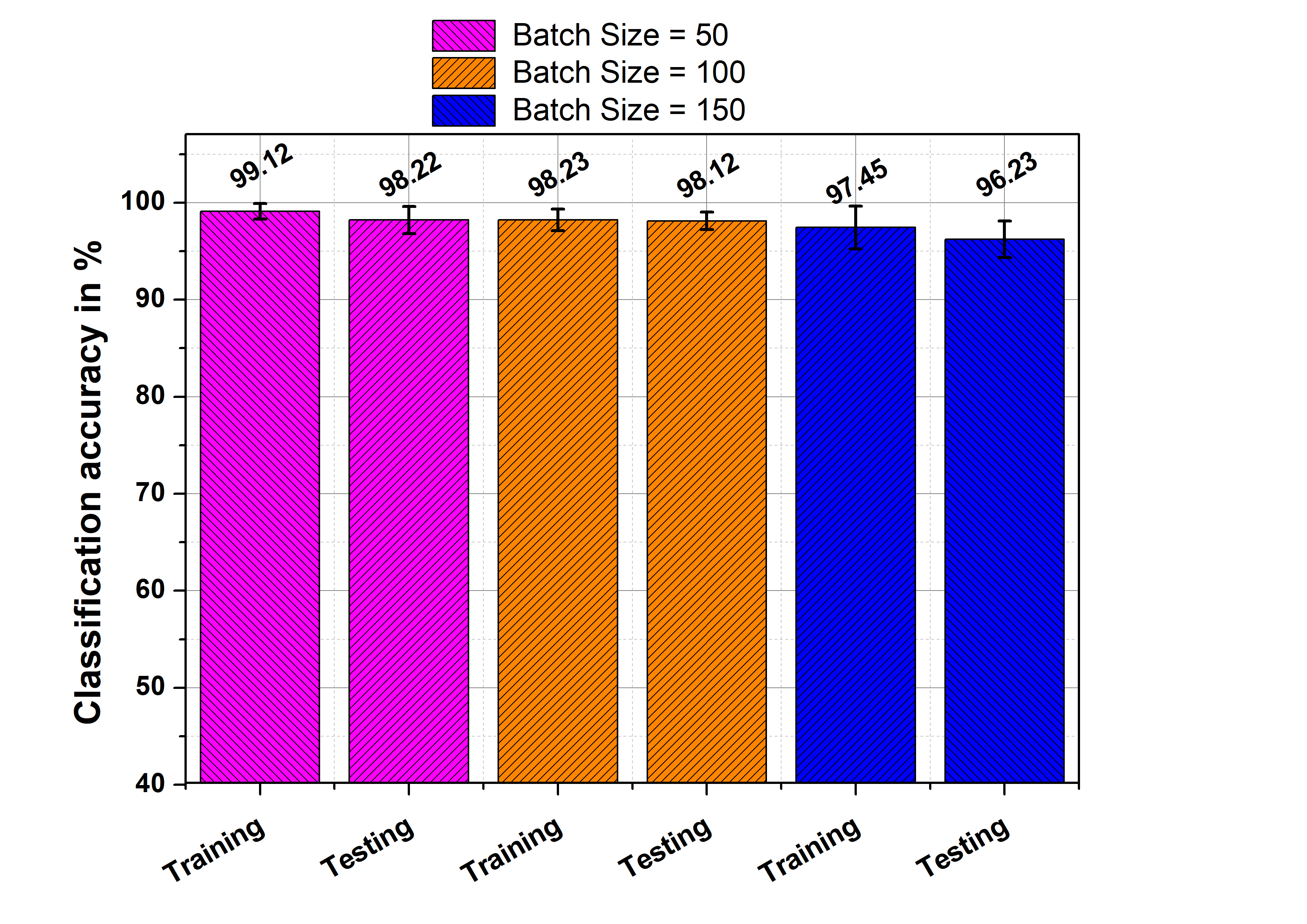}}
		\caption{Classification performance: (a) data-set-1, Batch size=50, (b) data-set-1, Batch size=100, (c) data-set-1, Batch size=150, (d) data-set-2, Batch size=50, (e) data-set-2, Batch size=100, (f) data-set-2, Batch size=150, (g) data-set-3, Batch size=50,(h) data-set-3, Batch size=100,(i) data-set-3, Batch size=150 \\}
		\label{cls1}	
	\end{figure*}

	\begin{table*}[htbp]
		\centering
		\caption{Classification accuracy comparison with machine learning and deep learning techniques}
		\begin{tabular}{llllll}
			\hline \hline
			&                                 &                                 & TD fearures                     &                                &                                   \\ \hline 
			\multicolumn{1}{c}{\begin{tabular}[c]{@{}c@{}}Datasets $\rightarrow$\\ $\downarrow$ Classifiers  \end{tabular}} & \multicolumn{1}{c}{NinaPro DB1} & \multicolumn{1}{c}{NinaPro DB2} & \multicolumn{1}{c}{NinaPro DB3} & \multicolumn{1}{c}{DB4(Force)} & \multicolumn{1}{c}{DB5(Position)} \\ \hline
			k-NN                                                                               & 60.34$\pm$4.3                           & 65.55$\pm$3.2                           & 33.46 $\pm$2.3                           & 85.65 $\pm$4.1                        & 88.90 $\pm$3.8                             \\
			SVM                                                                                & 64.76$\pm$5.6                            & 69.32$\pm$4.9                            & 40.46 $\pm$5.6                           & 88.43$\pm$2.9                       & 86.32$\pm$3.3                             \\
			RF                                                                                 & 69.44$\pm$2.8                           & 71.67$\pm$4.5                           & 52.16$\pm$4.2                          & 91.12$\pm$2.1                           & 93.12$\pm$3.4                              \\
			LDA                                                                                & 62.54$\pm$6.1                            & 64.84$\pm$5.4                            & 35.62$\pm$6.9                          & 85.34$\pm$5.2                           & 85.45$\pm$3.2                              \\ \hline
			&                                 &                                 & TFD features                    &                                &                                   \\ \hline
			k-NN                                                                               & 64.31$\pm$5.8                            & 67.12$\pm$4.3                           & 43.31$\pm$7.1                            & 87.42$\pm$4.7                           & 89.12$\pm$3.4                              \\
			SVM                                                                                & 69.76$\pm$4.1                           & 74.12$\pm$5.9                           & 51.89$\pm$6.8                          & 89.54$\pm$2.8                          & 87.32$\pm$4.2                             \\
			RF                                                                                 & 70.43$\pm$5.1                           & 74.69$\pm$5.6                           & 55.56$\pm$7.2                           & 93.11$\pm$1.8                          & 95.46$\pm$2.6                             \\
			LDA                                                                                & 64.57$\pm$6.4                           & 67.90$\pm$5.5                           & 40.62$\pm$7.4                           & 87.34$\pm$4.5                          & 88.66$\pm$3.4                             \\ \hline
			CNN  (2D Spectrogram)                                                              & 69.33                           & 73.12                           & 66.31                           & 93.12                          & 96.43                             \\ \hline
			Proposed framework                                                                 & 91.11                           & 89.45                           & 81.67                           & 95.67                          & 99.12 \\
			\hline \hline                         
		\end{tabular} 
		\footnotesize{ \\ k-nearest neighbors(k-NN), support vector machine(SVM), random forest (RF), linear discriminant analysis (LDA), convolutional neural network (CNN)}
		\label{ml}
	\end{table*}

	\begin{table*}[]
		\centering
		\caption{Comparison with previous work}
		\begin{tabular}{llllll}
			\hline \hline
			Author                                                                         & \begin{tabular}[c]{@{}l@{}}Subject\\  type\end{tabular}             & Dataset                                                                                                                           & Classes                                                      & Technique                                                                      & \begin{tabular}[c]{@{}l@{}}Classification \\ accuracy in \%\end{tabular}      \\ \hline
			\begin{tabular}[c]{@{}l@{}}Atzori et al. \cite{atzori2016deep} \\        (2016)\end{tabular}         & \begin{tabular}[c]{@{}l@{}}Healthy\\    and\\ Amputees\end{tabular} & \begin{tabular}[c]{@{}l@{}}NinaPro DB1\\ NinaPro DB2\\ NinaPro DB3\end{tabular}                                                   & \begin{tabular}[c]{@{}l@{}}50\\ 49\\ 49\end{tabular}         & \begin{tabular}[c]{@{}l@{}}All features\\ (TD, HIST, RMS,\\ mDWT)\end{tabular} & \begin{tabular}[c]{@{}l@{}}75.32 (RF)\\ 75.27 (RF)\\ 46.27 (SVM)\end{tabular} \\ \hline
			\begin{tabular}[c]{@{}l@{}}Zhai et al. \cite{zhai2017self}\\        (2017)\end{tabular}            & Healthy                                                             & NinaPro DB2                                                                                                                       & 49                                                           & CNN                                                                            & 78.71                                                                         \\ \hline
			\begin{tabular}[c]{@{}l@{}}Cotˆ e-Allard et al. \cite{cote2019deep} \\         (2019)\end{tabular} & Healthy                                                             & \begin{tabular}[c]{@{}l@{}}Hand/wrist\\ NinaPro DB5\end{tabular}                                                                  & \begin{tabular}[c]{@{}l@{}}7\\ 18\end{tabular}               & \begin{tabular}[c]{@{}l@{}}CNN\\ (Transfer learning)\end{tabular}              & \begin{tabular}[c]{@{}l@{}}98.31\\ 68.98\end{tabular}                         \\ \hline
			\begin{tabular}[c]{@{}l@{}}Zhang et al. \cite{zhang2019real}\\         (2019)\end{tabular}         & Healthy                                                             & (Dynamic+Air)                                                                                                                     & 10                                                           & LSTM                                                                           & 89.28                                                                         \\ \hline
			\begin{tabular}[c]{@{}l@{}}wei et al. \cite{wei2019surface} \\         (2019)\end{tabular} & Healthy                                                             & \begin{tabular}[c]{@{}l@{}}NinaPro DB1\\ NinaPro DB2\end{tabular}                                                                  & \begin{tabular}[c]{@{}l@{}}52\\50\end{tabular}               & \begin{tabular}[c]{@{}l@{}}CNN\\ (Multi-view (MV))\end{tabular}              & \begin{tabular}[c]{@{}l@{}}88.20\\ 83.70\end{tabular}\\  \hline
			Proposed                                                                       & \begin{tabular}[c]{@{}l@{}}Healthy\\    and\\ Amputees\end{tabular} & \begin{tabular}[c]{@{}l@{}}NinaPro DB1\\ NinaPro DB2\\ NinaPro DB3\\ Grasping (force variation)\\ Position variation\end{tabular} & \begin{tabular}[c]{@{}l@{}}53\\ 49\\ 49\\ 6\\ 8\end{tabular} & DLPR                                                                      & \begin{tabular}[c]{@{}l@{}}91.11\\ 89.45\\ 81.67\\ 95.67\\ 99.12\end{tabular} \\ \hline \hline
		\end{tabular}

		\label{lit}
	\end{table*}
	
	\section{RESULTS}
	The performance of the proposed DLPR based framework using DNN is validated using five different data-sets. The window size of 100 samples with a shifting of 10 samples has been selected when the proposed framework is used for NinaPro DB1 and a window size of 300 samples with shifting 50 samples has been considered for other remaining data-sets. The first three data-sets (NinaPro DB1, NinaPro DB2, NinaPro DB3) have been tested using three different batch sizes (50, 100, and 150). The training and testing classification accuracies have been shown in Fig. \ref{cls1} (a), (d), and (g)  for 50 epochs. The classification accuracies for the NinaPro DB1 of 91.11\%, 90.95\%, and 88.45\%  have been observed for the batch size of 50, 100, and 150 respectively, as shown in Fig. \ref{cls1} (a), (b), and (c) . For the NinaPro DB2, the classification accuracy of 88.43\%, 89.45\%, and 87.49\% have been seen for the three batch sizes, presented in Fig. \ref{cls1} (d), (e), and (f). The classification accuracy of 79.81\%, 81.67\&, and 80.44\% have been exhibited when NinaPro DB3 has been taken into consideration in Fig.\ref{cls1} (g), (h), and (i)  for the same batch sizes. Pattern classification of data-set 4  shows 95.67\% whereas pattern recognition rate for data-set 5 increased to 99.12\%  after 20 epochs as denoted in Fig. \ref{cls1} (a) and (b) respectively.    
	\subsection{Classification performance using time-domain(TD)  features} In this piece of work, the classification performance of all five data-sets has been calculated using TD features. Total four well-known classification algorithms such as k-NN (k-Nearest Neighbors), SVM (support vector machine) using the quadratic kernel, RF (random forest), and LDA (linear discriminant analysis) have been considered. The window length of 200 ms with 75 ms increment is fixed for the feature extraction.\par 
	The highest classification accuracies of 69.44$\pm$2.8\%, 71.67$\pm$4.5\%, 52.16$\pm$4.2\%, 91.12$\pm$2.1\%, and 93.12$\pm$3.4\% have been achieved for NinaPro DB1, NinaPro DB2 ,NinaPro DB3, data-set-4 and dataset-5 respectively using RF classification technique. The lowest pattern classification performance has been achieved by the LDA classifier as summarized in Table \ref{ml}.

	\subsection{Classification perfromance using time-frequency domain (TFD) features} 
	The classification performance using classical TFD features has been tested using all the classifiers and result is denoted in Table \ref{ml}. Highest classification accuracies of 70.43$\pm$5.1\%, 74.69$\pm$5.6\%, 55.56$\pm$7.2\%, 93.11$\pm$1.8\%, and 95.46$\pm$2.6 have been exhibited for all the considered data-sets receptively. The lowest classification accuracy for NinaPro DB1 and NinaPro DB2 has been observed when k-NN classier is employed due to highly homogeneous activities. In the rest of the data-sets, LDA shows the lowest classification accuracy. 
	
	\subsection{Classification performance using CNN (spectrogram images)}
	Classification performance using STFT based spectrogram images and convolution neural network is shown in Table \ref{ml}. The pattern recognition rate of 69.33\%, 73.12\%, 66.31\%, 93.12\%, and 96.43\% have been achieved for the NinaPro DB1,  NinaPro DB2, NinaPro DB3, data-set 4, and data-set 5 respectively. The pattern classification performance is boosted over TD and TFD methods.

	\section{Discussion}
\begin{figure}[htbp]
		\centering
		\includegraphics[width=0.35\textwidth,height=0.25\textheight]{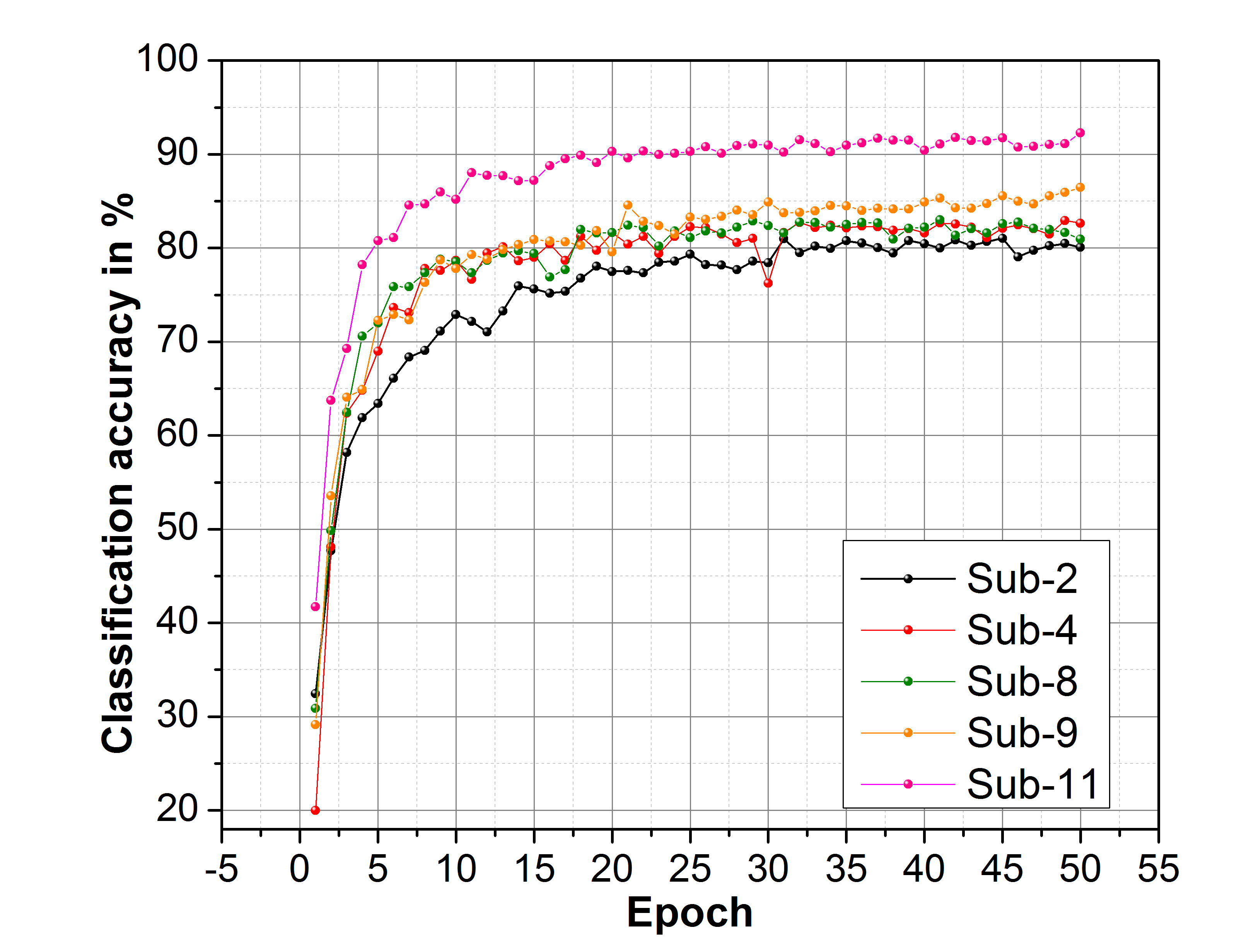}
		\caption{NinaPro DB3 classification accuracies for different amputees}
		\label{cls}	
	\end{figure}

	The proposed DLPR method exhibits significant improvement in classification performance with respect to classical machine learning techniques and with spectrogram images based on CNN architecture. In this paper, NinaPro DB3 has been considered for evaluating pattern recognition performance on amputated subjects as shown in Fig. \ref{cls}. All the amputated subjects have different physical properties in terms of DASH score, remaining forearm, and phantom limb sensation. Subject 11, who had lost his limb due to cancer and having DASH score 12.5 achieved a classification accuracy of 92.26\%. Subject 2 (DASH=15.18) and subject 8 (DASH=33.33) have achieved the lowest classification accuracy of 80.34\% and 80.92\% respectively. Subject 9 has a DASH score of 3.3 and has achieved 86.45\% classification accuracy. This shows that pattern recognition performance has been boosted remarkably when NinaPro DB3 has been taken into consideration. Furthermore, this work is compared with literature and is also summarized in Table \ref{lit}.  In the research \cite{atzori2016deep}, three data-sets were considered and classification accuracies of 75.32\% (RF), 75.27\% (RF), and 46.27\% (SVM) were achieved for NinaPro DB1, NinaPro DB2, and NinaPro DB3  respectively. Paper \cite{zhai2017self} reported  78.71\%  classification accuracy for the NinaPro DB2 (healthy subjects and 49 classes) using CNN. Total 2 data-sets Hand/writ and NinaPro DB5 (18 grasping activities) have been considered in the research \cite{cote2019deep}. The classification accuracy of 98.31\% and 68.98\% have been achieved for these 2 data-sets respectively. Total 10 upper limb activities (Dynamic and Air) were classified using LSTM (low short term memory) network and classification accuracy of 89.28\% was obtained. Data-sets (NinaPro DB1 and NinaPro DB2) have been used and a multi-view deep learning framework was employed. In this research, the pattern recognition rate 88.20\% and 83.70\% were achieved for NinaPro DB1 and NinaPro DB3 respectively. However, data with force variation and position variation have not been considered. Proposed DLPR architecture has been examined for 5 different types of scientific and benchmark data-sets and the pattern classification rate is significantly boosted for all healthy and amputated subjects. At the initial stage, the raw EMG signal has been passed through two parameters (MZP and MPP) which reduces data size for the training with minimal loss of neural information. These parameters contain information regarding time as well as frequency present and this helps to train a highly accurate classification model. The proposed model shows robust performance by showing significant enhancement in classification accuracy in all kinds of data-sets. \par
	 Furthermore, training time for different data-set have been evaluated as represented in Table \ref{time}. This demonstrate that this model only takes few minutes time to get the high classification accuracy.

\begin{table}[htbp]
\centering
\caption{Training time of proposed model for different data-sets}
\begin{tabular}{lll}
\hline \hline
S. No & Data-set & \begin{tabular}[c]{@{}l@{}}Training time \\ (50 epochs)\end{tabular} \\ \hline
1.    & DB1      & $\approx458$ sec                                                              \\
2.    & DB2      & $\approx523$ sec                                                              \\
3.    & DB3      & $\approx523$ sec                                                              \\
4.    & DB4      & $\approx210$ sec                                                              \\
5.    & DB5      & $\approx459$ sec                                                              \\ \hline \hline
\end{tabular}
\label{time}
\end{table}

	\section{conclusion}
The proposed DLPR-based system is a robust, fast, and precise deep learning-based framework for classifying diverse gripping and functional activities for upper limb prosthesis applications. The investigation is carried out on five distinct types of standard and benchmark data-sets, comprising healthy and amputee people. This proposed approach is used as a pre-processing step in the primary phase to make input information-rich and stationary. With current equivalents, considerable functional development has been noticed in terms of classification accuracy of up to 25\% and time complexity is decreased significantly. Classification performance of amputees with a high DASH score is also improved. In future work, this techniques will be translated into DSP processor for real-time EMG pattern classification.

	\section{Acknowledgement}
	The authors would like to acknowledge the support of the Ministry of Human Resource and Development, India, Bionics and Intelligence Lab. Jaipur, India, and Dr. Rami N Khushaba  and Dr. Ali H. Al-Timemy for data-sets 4 and 5.
	
	\bibliographystyle{IEEEtran}
	\bibliography{DLPR}

\begin{IEEEbiography}
		[{\includegraphics[width=1in,height=1.00in,clip,keepaspectratio]{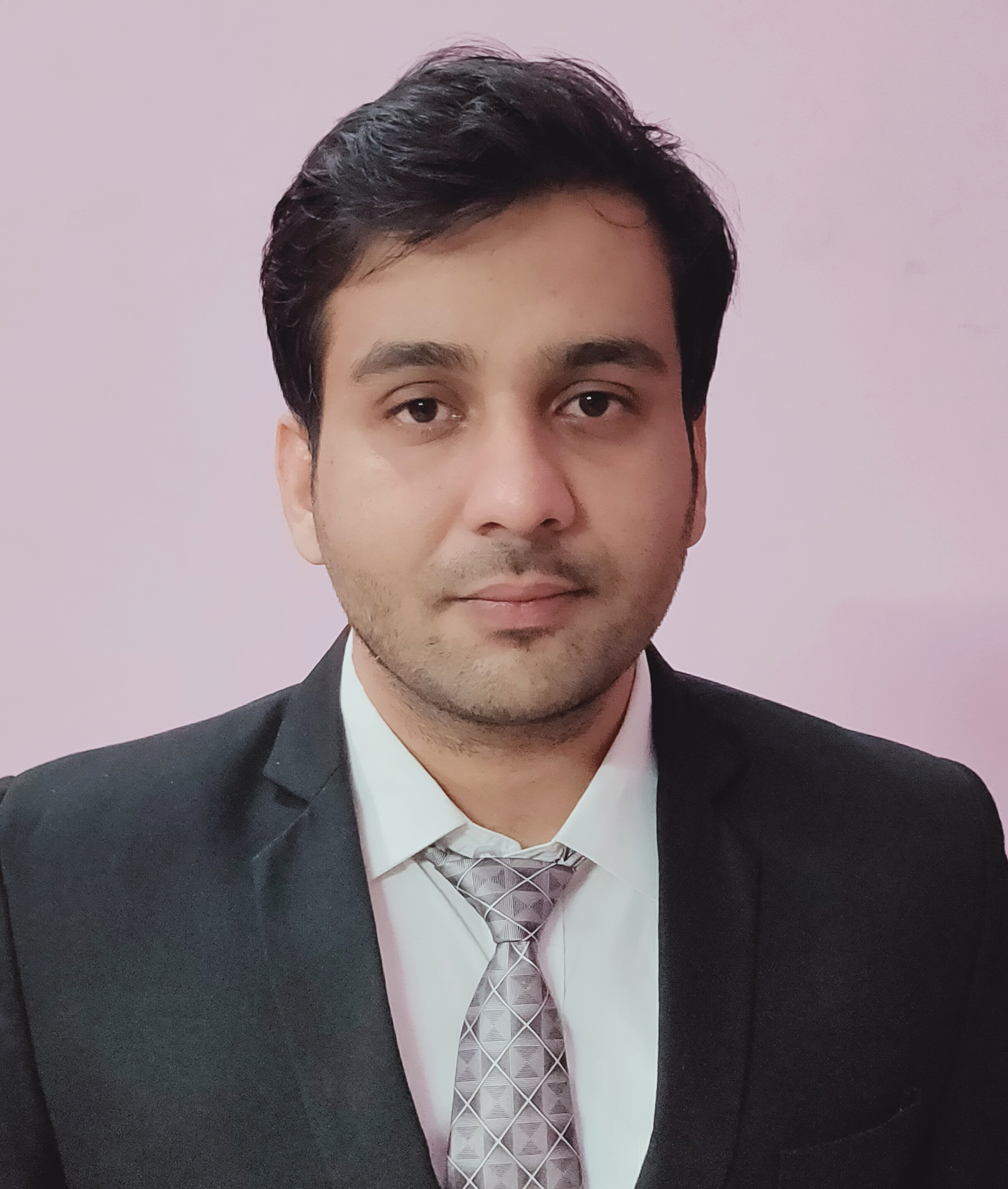}}]
		{Sidharth Pancholi} received the Masters from Thapar University, Patiala, India, in 2016. He has completed his PhD work with Malaviya National Institute of Technology, Jaipur, India. Currently, he is working as a Research Associate with the Indian Institute of Technology, Delhi, India. He has worked as a reviewer of technical journals such as IEEE Transactions/ journals and served as Technical Programme Committee member for IEEE conferences. He also received CSIR Travel fellowship and CSSTDS Travel fellowship to attend IEEE Conference EMBC 2019. He has two years of industry experience. His current research interests include biomedical signal processing, neural rehabilitation, Brain computer interface, Human-machine interface, prosthetic device development, and embedded systems.
	\end{IEEEbiography}

	\begin{IEEEbiography}
		[{\includegraphics[width=1in,height=1.00in,clip,keepaspectratio]{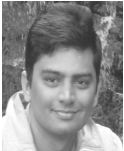}}]
		{Amit M Joshi} (M'08) has completed his M.Tech (by research) in 2009 and obtained Doctoral of Philosophy degree (Ph.D) from National Institute of Technology, Surat in August,2015. He is currently working as Assistant Professor at National Institute of Technology, Jaipur since July,2013. His area of specialization is Biomedical signal processing, Smart healthcare, VLSI DSP Systems and embedded system design. He has published six book chapters and also published 50+ research articles in excellent peer reviewed international journals/conferences. He has worked as a reviewer of technical journals such as IEEE Transactions, Springer, Elsevier and also served as Technical Programme Committee member for IEEE conferences. He also received UGC Travel fellowship , SERB DST Travel grant  and CSIR Travel fellowship to attend IEEE Conferences in VLSI and Embedded System. He has served session chair at various IEEE Conferences like TENCON -2016, iSES-2018, ICCIC-14 etc.
	\end{IEEEbiography}
	
	\begin{IEEEbiography}
		[{\includegraphics[width=1in,height=1.00in,clip,keepaspectratio]{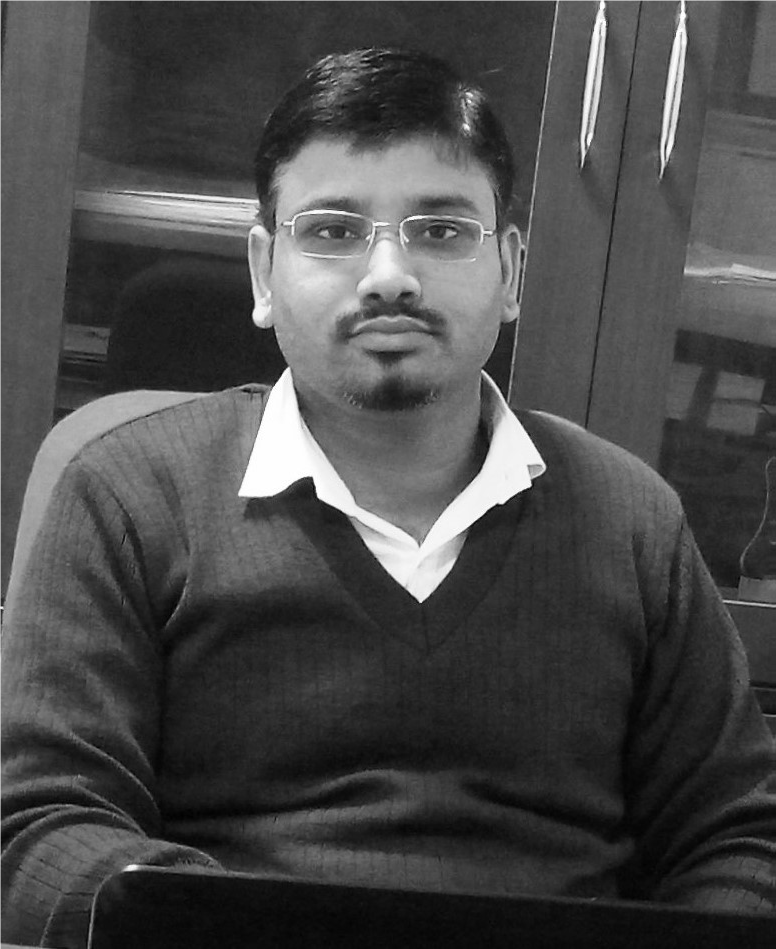}}]
		{Deepak Joshi} (M'08) received the Ph.D. degree in biomedical engineering from IIT Delhi, New Delhi, India, in 2012. He worked at various places, including the National University of Singapore, Singapore, Newcastle University, Newcastle upon Tyne, U.K., and the University of Oregon, Eugene, OR, USA. He is currently a Faculty Member in biomedical engineering with IIT Delhi and the All India Institute of Medical Sciences Delhi, New Delhi. He has been working in the areas of neuroprosthetics and neurorehabilitation for last nearly 14 years. His current research work combines experimental and computational techniques to understand the neural correlates during balancing and seamless transitions during walking. Besides that, he is also actively engaged in projects related to the development of wearable devices for applications specific to the diagnosis of neuromuscular disorders, assistive devices for the elderly and disabled, and biofeedback for rehabilitation in stroke patients.
	\end{IEEEbiography}

	\vspace{-1.2cm}

\end{document}